\documentclass{raa_twocolumn}          

\usepackage{graphicx,times}             
\usepackage{natbib}
\usepackage{amssymb,amsmath}
\bibpunct{(}{)}{;}{a}{}{,}
\usepackage[pagebackref=true]{hyperref}

\begin{document}

  \title{SVOM Science User Support Services at Chinese Science Center}

   \volnopage{Vol.0 (202x) No.0, 000--000}      
   \setcounter{page}{1}          

   \author{Xu-hui Han 
      \inst{1,2}
   \and Li-ping Xin
      \inst{1,2}
   \and Jing Wang
      \inst{1}
   \and Yu-jie Xiao
      \inst{1}
   \and Pin-pin Zhang
      \inst{1}
   \and Ruo-song Zhang
      \inst{1}
   \and Hong-bo Cai
      \inst{1}
   \and Hui-jun Chen
      \inst{3,1}
   \and Jin-song Deng
      \inst{1,2}
   \and Wen-long Dong
      \inst{1}
   \and Lei Huang
      \inst{1}
   \and Lin Lan
      \inst{1}
   \and Hua-li Li
      \inst{1}
   \and Guang-wei Li
      \inst{1,2}
   \and Xiao-meng Lu
      \inst{1}
   \and Yu-lei Qiu
      \inst{1}
   \and Chao Wu
      \inst{1,2}
   \and Wen-jin Xie
      \inst{1}      
   \and Da-wei Xu
      \inst{1,2} 
   \and Jing-ran Xu
      \inst{1} 
   \and Yang Xu
      \inst{1} 
   \and Zhu-heng Yao
      \inst{1} 
   \and Mo Zhang
      \inst{1} 
   \and Xue-ying Zhao
      \inst{1} 
   \and Wei-kang Zheng
      \inst{4} 
   \and Ya-tong Zheng
      \inst{1}
   \and Xiao-xiao Zhou
      \inst{5}
   \and Jian-yan Wei
      \inst{1}
   }

   \institute{National Astronomical Observatories, Chinese Academy of Sciences, Beijing, 100101, People’s Republic of China. {\it hxh@nao.cas.cn} \\
        \and
    School of Astronomy and Space Science, University of Chinese Academy of Sciences, Beijing, 101408, People’s Republic of China.\\
        \and
    Department of Computer Science and Technology, School of Computer 
    and Information Technology Cangzhou Jiaotong College, Hebei, 061199, 
    People’s Republic of China.
        \and
    Department of Astronomy, University of California, Berkeley, CA 94720-3411, USA.
        \and
    College of Information Engineering, Fuzhou Polytechnic, Fuzhou, Fujian, 350108, People's Republic of China.
\vs\no
   {\small Received 202x month day; accepted 202x month day}}

\abstract{The Chinese-French \textit{SVOM} (Space-based Multi-band Astronomical Variable Objects Monitor) mission is dedicated to the study of gamma-ray bursts (GRBs) from the distant universe. A key component of the \textit{SVOM} Chinese Ground Segment, the Science User Support Services (SUSS) provides comprehensive support for the mission's scientific operations. SUSS consists of two integral pillars: a suite of specialized software tools that automate key workflows, and a dedicated User Support Team that delivers expert-led, human services. These human-delivered services include operational coordination across telescope networks, direct technical assistance to astronomers, user training, and proactive problem-solving throughout the observation lifecycle.
This paper focuses on the organization of \textit{SVOM} scientific operations and the role of SUSS in facilitating these tasks. We provide a detailed description of the SUSS software architecture and its functionalities, encompassing the General Platform, the Burst Advocate (BA) support tools for GRB counterpart identification, the Target of Opportunity (ToO) support tools, and the General Program (GP) support tools. The structure and services provided by the user support team at the Chinese Science Center (CSC) are also elaborated. Furthermore, we evaluate the performance of SUSS during its first operational year, assessing its effectiveness in fulfilling user requirements. The evaluation offers valuable insights to guide future user support strategies and software enhancements, ultimately enabling better service for the \textit{SVOM} scientific community.
}

   \authorrunning{X. Han, L. Xin, J. Wang, et al.}            
   \titlerunning{SVOM Science User Support Services at CSC}  

   \maketitle

%
%
\section{Introduction}           
\label{sect:intro}

The \textit{SVOM} (Space-based Multi-band Astronomical Variable Objects Monitor) mission is a joint Chinese-French satellite project designed to detect, localize, and study gamma-ray bursts (GRBs) and other high-energy transient phenomena. Its primary scientific goals are to ensure a continuous supply of well-localized explosive transients for multi-wavelength follow-up studies and to provide the most comprehensive characterization of the detected events. To achieve these objectives, \textit{SVOM} employs an innovative concept combining space-borne and ground-based instruments: the satellite carries four complementary instruments—including the wide-field hard X-ray telescope ECLAIRs for GRB detection and localization, and narrow-field telescopes for afterglow observations—while a dedicated ground segment, including the GWAC wide-field camera and robotic follow-up telescopes, enables rapid multi-wavelength observations from the prompt emission to the late afterglow phases.

The scientific operations of the \textit{SVOM} mission are organized into three main programs, each with distinct observational requirements. The Core Program (CP) consists of observations triggered automatically by ECLAIRs alerts, focusing on gamma-ray bursts (GRBs) and other high-energy transient sources. This program requires rapid validation and follow-up coordination, managed through dedicated burst advocate tools. The Target of Opportunity Program (ToO) enables ground-commanded observations with swift reaction times—ranging from minutes to hours—allowing the mission to respond to transients detected by other observatories. This time-critical process demands efficient proposal submission, review, and scheduling capabilities. Finally, the General Program (GP) covers all pre-planned observations conducted during quiescent periods, with targets defined annually by the scientific community, requiring comprehensive tools for proposal management, review, and long-term observation planning.

The \textit{SVOM} Chinese Ground Segment undertakes several important functions for organizing and supporting these scientific operations (\citet{Liu2026} and \citet{Huang2026}). For the scientific management of the CP, ToO, and GP, the \textit{SVOM} Chinese Science Center (CSC) has developed the Science User Support System (SUSS). The SUSS website (\footnote{\url{https://www.svom.cn}}) hosts three main scientific user support tools: CSC BA Tools (for CP), ToO Tools, and GP Tools.

\begin{figure}
	\centering 
    \includegraphics[page=1,width=1\linewidth]
	{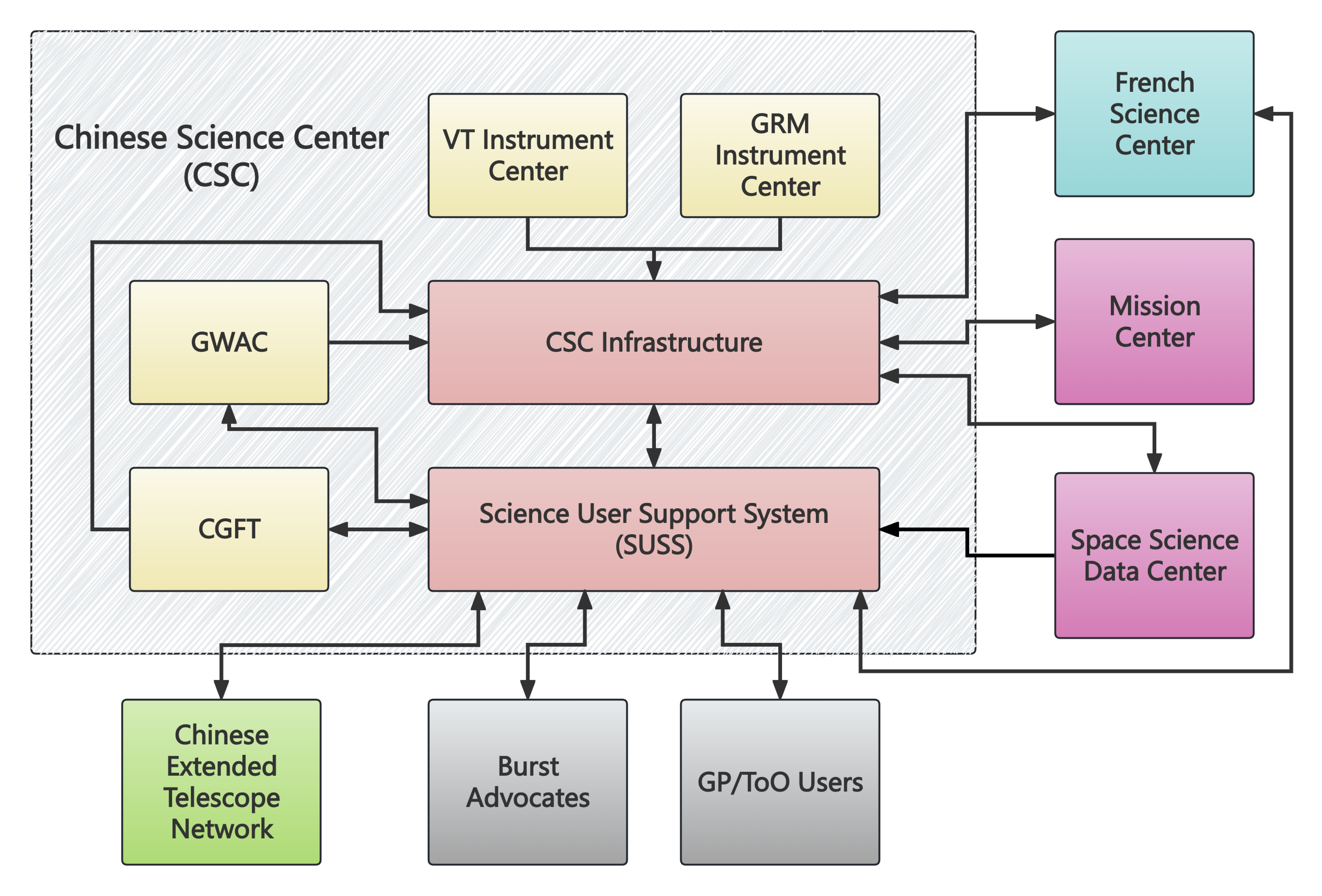}	
	\caption{The CSC infrastructure and SUSS, integral components of the CSC.} 
	\label{fig:organization_of_suss}
\end{figure}

The overall structure of the \textit{SVOM} Chinese Ground Segment is shown in Figure \ref{fig:organization_of_suss}. It demonstrates that the CSC infrastructure and SUSS are integral components of the CSC. The infrastructure handles the processing and transmission of information 
and scientific products \citep{Huang2026}, 
while SUSS supports most scientific activities within the CSC.

SUSS assists \textit{SVOM} scientific users in completing tasks related to the organization of scientific activities in three main programs: CP, GP, and ToO \citep{Cordier2026}. These include BA duty management, GRB alert notifications, GRB follow-up observation proposals, submission of ToO and GP proposals, proposal review, generation and submission of observation requests and target lists, feedback on observation status, and querying and downloading observation data. Simultaneously, SUSS provides users with information on the status of \textit{SVOM} satellite payloads, ground telescopes, and observations.

SUSS integrates the business logic, user management, real-time notification services, underlying databases, and front-end/back-end components of the three scientific support tools (CSC BA Tools, ToO Tools, and GP Tools) within a unified technical framework. It provides services to users via a web interface.

The rest of this paper is organized as follows. Section \ref{architecture_suss} describes the architecture of SUSS, including the General Platform and the three main support tools. Section \ref{science_operation} details the arrangement of the science operations for the CP, ToO, and GP. Section \ref{suss_performance} presents the performance of SUSS in its first operation year. Finally, Section \ref{conclusion} concludes the paper.

\section{Architecture of SUSS}
\label{architecture_suss}

The SUSS is architected as a unified system comprising four interrelated layers: the General Platform, which provides foundational services, and three specialized tool sets—CSC BA Tools, ToO Tools, and GP Tools—each tailored to the specific requirements of the \textit{SVOM} scientific programs introduced in Section \ref{sect:intro}.

The General Platform serves as the common backbone, offering shared functionalities such as user authentication, content management, data product handling, real-time push notifications, and a centralized portal for information dissemination. These core services are leveraged by all three specialized tools, ensuring consistency and reducing redundant development.

The CSC BA Tools are dedicated to the Core Program, supporting Burst Advocates in rapid GRB validation, follow-up coordination, and shift management. These tools integrate with the General Platform to receive real-time alerts and access GRB data products.

The ToO Tools address the time-critical demands of the Target of Opportunity Program, enabling swift proposal submission, review, and observation scheduling. They utilize the General Platform's notification services to alert ToO scientists of new proposals and its data management capabilities to organize ToO observation products.

The GP Tools support the General Program's annual cycle of proposal submission, peer review, time allocation, and long-term observation planning. They rely on the General Platform for user management, proposal archiving, and catalog generation.

Together, this layered architecture ensures that each scientific program benefits from both specialized functionalities and a robust, shared infrastructure, facilitating efficient and coordinated science operations across the \textit{SVOM} mission.

\begin{figure}
	\centering 
	\includegraphics[page=1,width=1\linewidth]{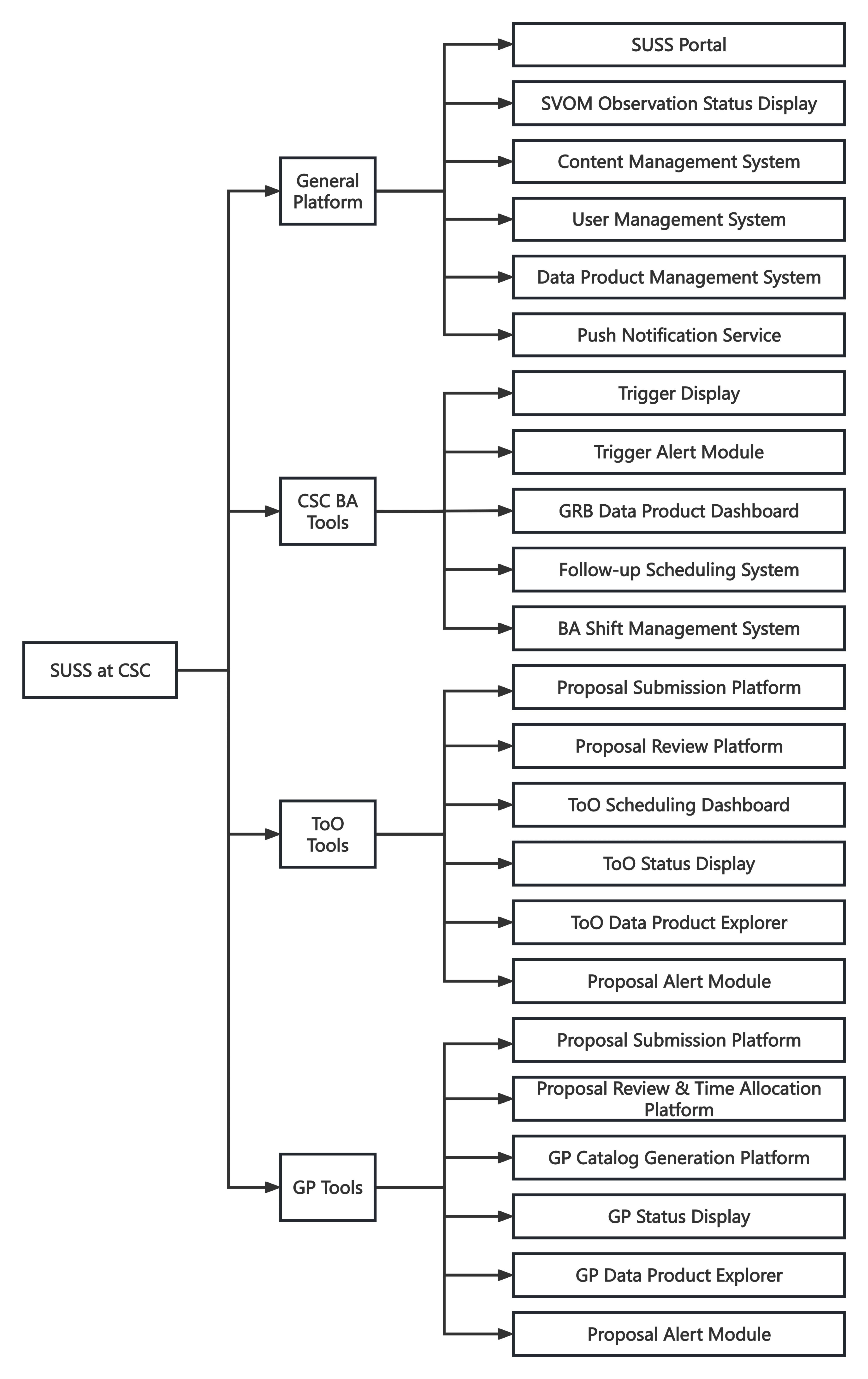}	
	\caption{The architecture of SUSS, which includes the General Platform, CSC BA Tools, ToO Tools, and GP Tools, is designed to integrate the distinct functions of each component.} 
	\label{fig:architecture_of_suss}
\end{figure}

The basic structure of SUSS is shown in Figure \ref{fig:architecture_of_suss}. This chapter elaborates 
on the functionalities of its various support tools.

\subsection{General Platform}

The SUSS General Platform provides common functions for the CSC BA Tools, ToO Tools, and GP Tools, including:

\subsubsection{SUSS Portal}

The SUSS Portal provides users with \textit{SVOM}-related news, GP proposal call announcements, user guides, various documents, and different \textit{SVOM} project links. It is an important channel for scientific users to obtain information related to \textit{SVOM} and its scientific activities. Articles are publicly displayed on the SUSS Portal according to their categories.

\subsubsection{\textit{SVOM} Observation Status Display}

SUSS provides users with the basic status of \textit{SVOM} scientific operations, enabling them to quickly understand \textit{SVOM}'s historical and current status. This assists users in making rapid decisions regarding GRB identification and ToO observation proposals.

\begin{figure*}
	\centering 
	\includegraphics[page=1,width=1\linewidth]{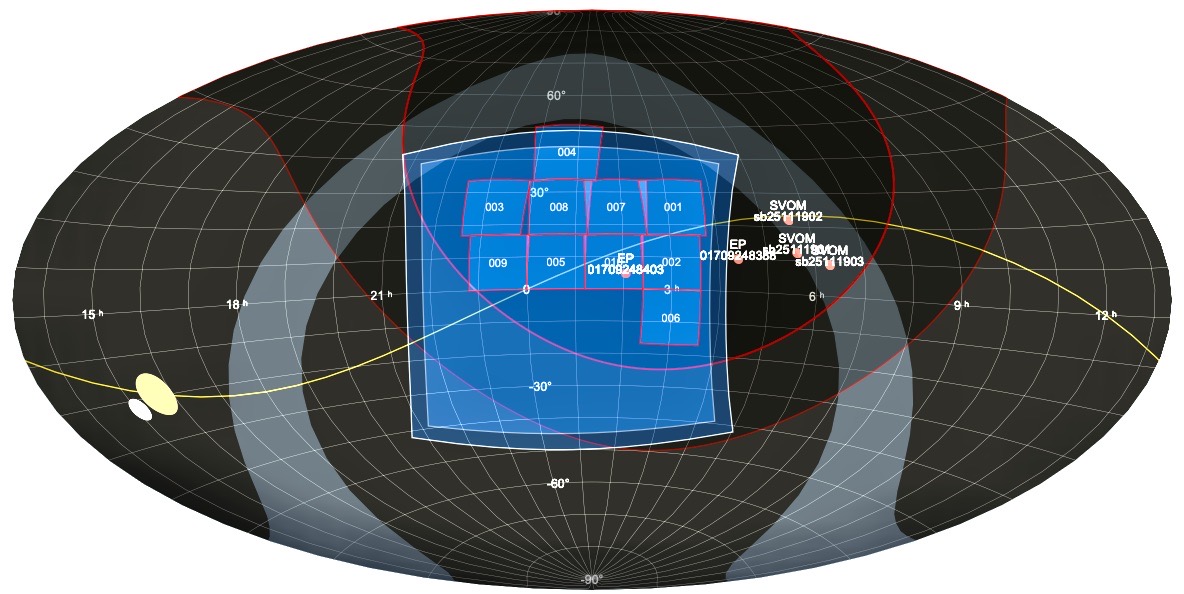}	
	\caption{The figure shows the coverage regions of \textit{SVOM}/Eclairs (blue areas) and GWAC-A (red boxes), with the alert positions and error regions of multiple events marked. Also displayed are the ecliptic plane (yellow curve), the positions of the Sun and the Moon, and the exclusion of low galactic latitude regions (gray areas).Two red curves delineate the area above a 
    25-degree elevation limit and a 0-degree elevation limit from the GWAC observation site. They demonstrate that within the observable sky for GWAC (elevation $>$ 25°)} 
	\label{fig:celestial_sphere_map}
\end{figure*}

SUSS provides a dynamic celestial sphere map (Figure \ref{fig:celestial_sphere_map}) displaying the alert positions and error regions of events detected by \textit{SVOM} and other astronomical satellites (\textit{Swift}  \citep{Gehrels04}, Fermi \citet{Atwood09}, \textit{Einstein Probe} (EP, \citet{Yuan22}), 
{\rm etc.}) at different times. The map also shows the monitored sky areas of \textit{SVOM}/Eclairs and GWAC (Ground-based Wide Angle Camera, \citet{Han2021} and \citet{Xin2026}). Users can quickly assess potential associations between events detected by different satellites and the monitored sky areas of \textit{SVOM}/Eclairs and GWAC.

\begin{figure*}
	\centering 
	\includegraphics[page=1,width=1\linewidth]{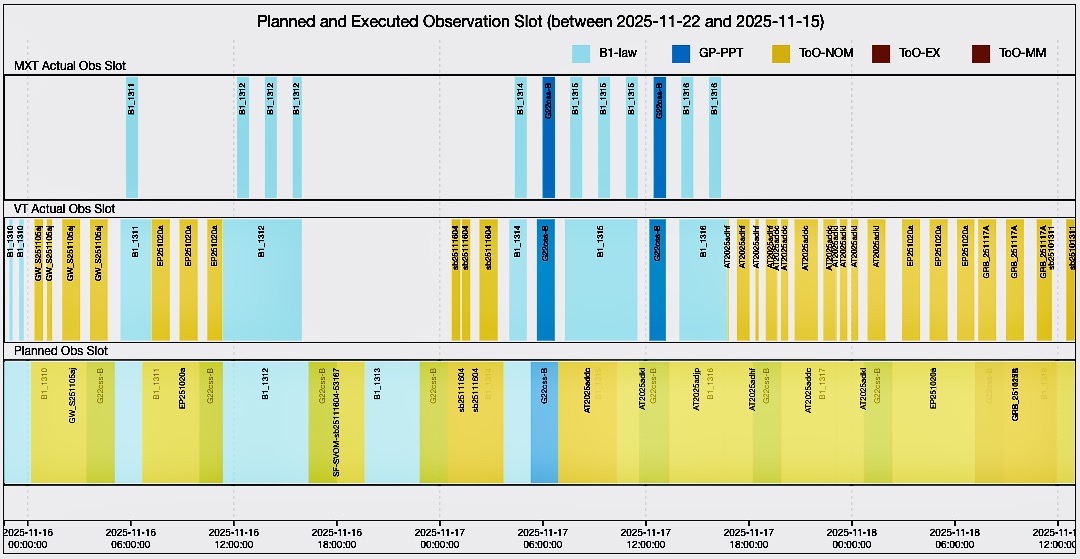}	
	\caption{The figure shows the ToO and GP observation plans (bottom panel), completion status of VT (middle panel) and MXT (top panel)at different times.} 
	\label{fig:gp_too_observation}
\end{figure*}

SUSS provides dynamic charts displaying the ToO and GP observation plans and completion status at different times, which is demonstrated in Figure \ref{fig:gp_too_observation}. Users can check ToO/GP observation targets, planned observation times, and completion status, enabling quick decisions on the necessity of ToO observation proposals and available observation time windows.

\begin{figure*}
	\centering 
	\includegraphics[page=1,width=1\linewidth]{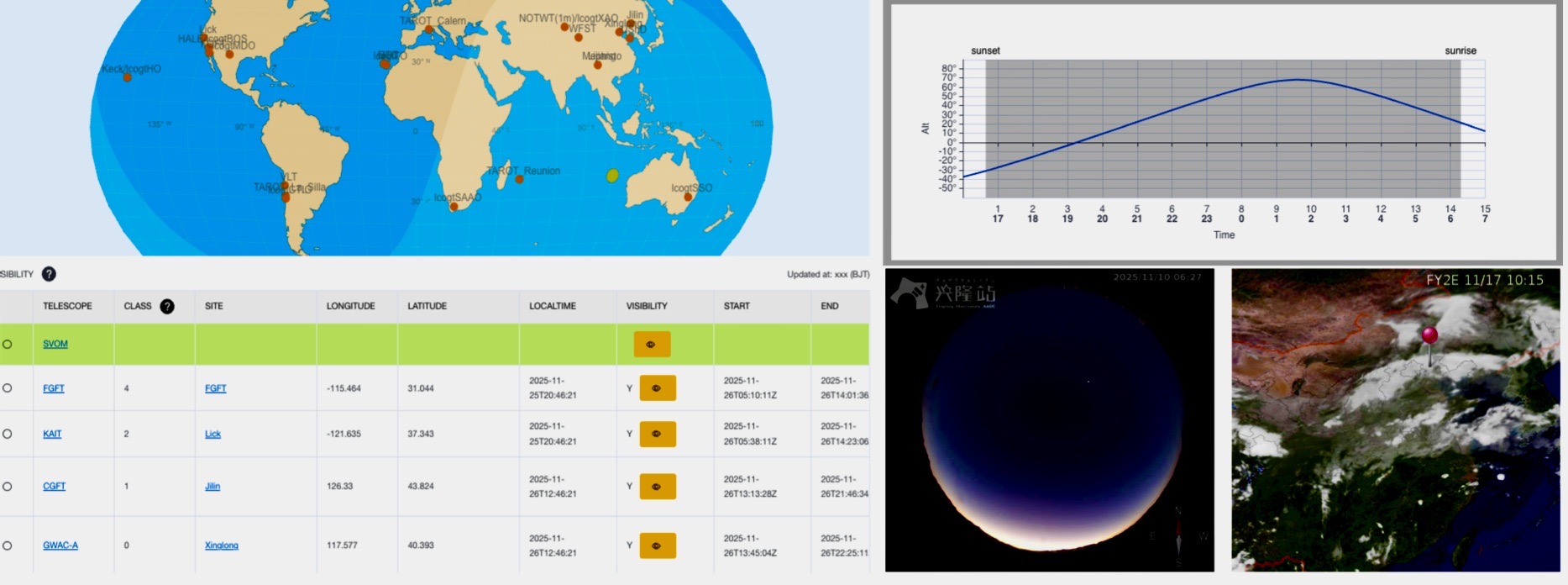}	
	\caption{The figure presents an overview of the \textit{SVOM} ground-based follow-up telescope system. The top-left panel shows the geographical distribution of the telescopes and their day/night periods; the bottom-left indicates their real-time availability status (green); the top-right displays the visibility of a target, expressed as its altitude; and the bottom-right provides local weather conditions and cloud coverage for each observatory.} 
	\label{fig:telescope_status}
\end{figure*}

SUSS provides dynamic charts and lists showing the geographical locations of different ground telescopes, day/night periods, telescope availability, and weather conditions at the telescope sites, which are shown in Figure \ref{fig:telescope_status}. Users can check target visibility and quickly select ground telescopes for GRB follow-up observations.

\subsubsection{Content Management System}

The Content Management System provides a backend web interface for administrators to create, edit, publish, sort, and retract the articles displayed on the portal.

\subsubsection{User Management System}

SUSS allows users to access all tools with a single account. The three support tools (CSC BA Tools, ToO Tools, and GP Tools) have different user types with varying permissions, including roles such as system administrator, PI, BA manager, BA, ToO scientist, GP manager, ToO/GP proposers, and GP Time Allocation Committee (TAC) members.

Many \textit{SVOM} scientific users have permissions to use multiple support tools. Therefore, SUSS provides a unified user management platform for system administrators and the administrators of each support tool to manage user identities, permissions, and accessible functional modules.

\subsubsection{Data Product Management System}

The Data Product Management System provides GRB and corresponding ToO/GP observation data products for \textit{SVOM} scientific users, such as BAs and ToO/GP proposers. The system collects scientific data products from both \textit{SVOM} space-based instruments and ground-based telescopes, while establishing correlations between data sources, observation tasks, and observation proposers to ensure precise data distribution.

\begin{figure*}
	\centering 
	\includegraphics[page=1,width=0.9\linewidth]{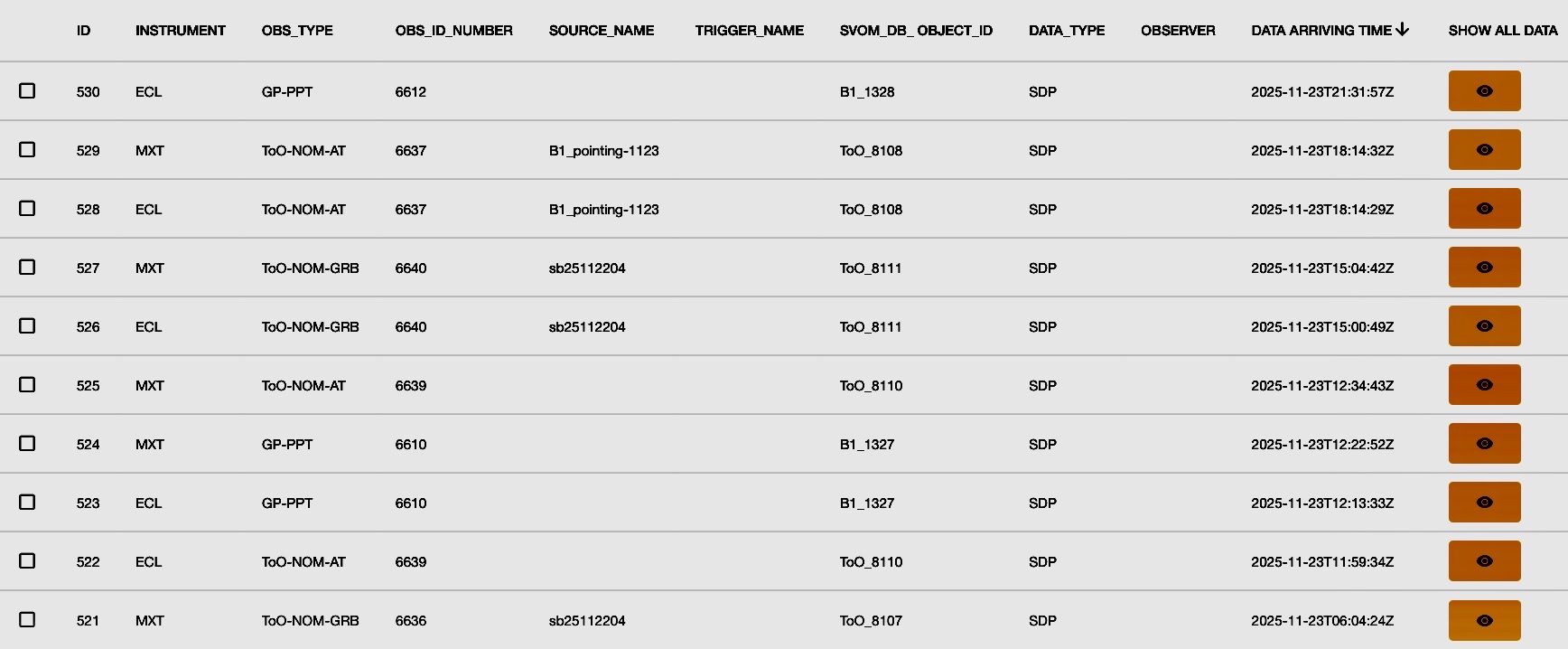}	
	\caption{A centralized interface in the Data Product Management 
    System is provided for accessing scientific data products, which are systematically associated with their corresponding data sources, observation tasks, and observation proposers.} 
	\label{fig:data_sdp}
\end{figure*}

\subsubsection{Push Notification Service}

The Push Notification Service offers underlying support for various instant messaging platforms (including Short Message Service (SMS), 
{\sc mattermost}(\footnote{\url{https://mattermost.com}}), 
{\sc enterprise wechat}(\footnote{\url{https://work.weixin.qq.com}}), 
{\sc dingtalk}(\footnote{\url{https://www.dingtalk.com}}), 
{\sc feishu}(\footnote{\url{https://www.feishu.cn}}), 
{\rm etc.}), tailored to the usage habits and environments of \textit{SVOM}'s international user base. Each support tool within SUSS can leverage this service to deliver critical real-time notifications, enabling users to promptly receive GRB alerts, follow-up observation requests, ToO/GP proposal status updates, and other key operational information.

\subsection{CSC BA Tools}

\subsubsection{Trigger Alert Module}

The Trigger Alert Module automatically delivers GRB notices to designated messaging platforms via the Push Notification Service. Through the tool's interface, BAs can select their preferred platform and subscribe to notices based on type.

\subsubsection{Trigger Display}

The CSC BA tools provide pages displaying GRB information, including the triggering satellite and instrument, event type, trigger time, position, error region, and signal-to-noise ratio. The tools also provide interfaces for users to record other relevant information about the GRB event, such as GRB type, redshift, extinction, and potential host galaxy. The interface is optimized for mobile devices, allowing BAs to efficiently monitor GRB alerts directly on their smartphones.

\subsubsection{GRB Data Product Dashboard}

The CSC BA tools organize scientific data products from space-based instruments and ground telescope detections by GRB event. BAs can view the scientific products for a specific GRB through the GRB Data Product Dashboard. The dashboard also visualizes quick-look scientific product images and light curves of optical counterpart candidates detected by the Visible Telescope (VT, \citep{Qiu2026}) and the Chinese Ground Follow-up Telescope (CGFT, \citep{Wu2026}), facilitating rapid identification of optical counterparts and assessment of their variability characteristics.

\subsubsection{Follow-up Scheduling System}

\begin{figure*}
	\centering 
	\includegraphics[page=1,width=1\linewidth]
    {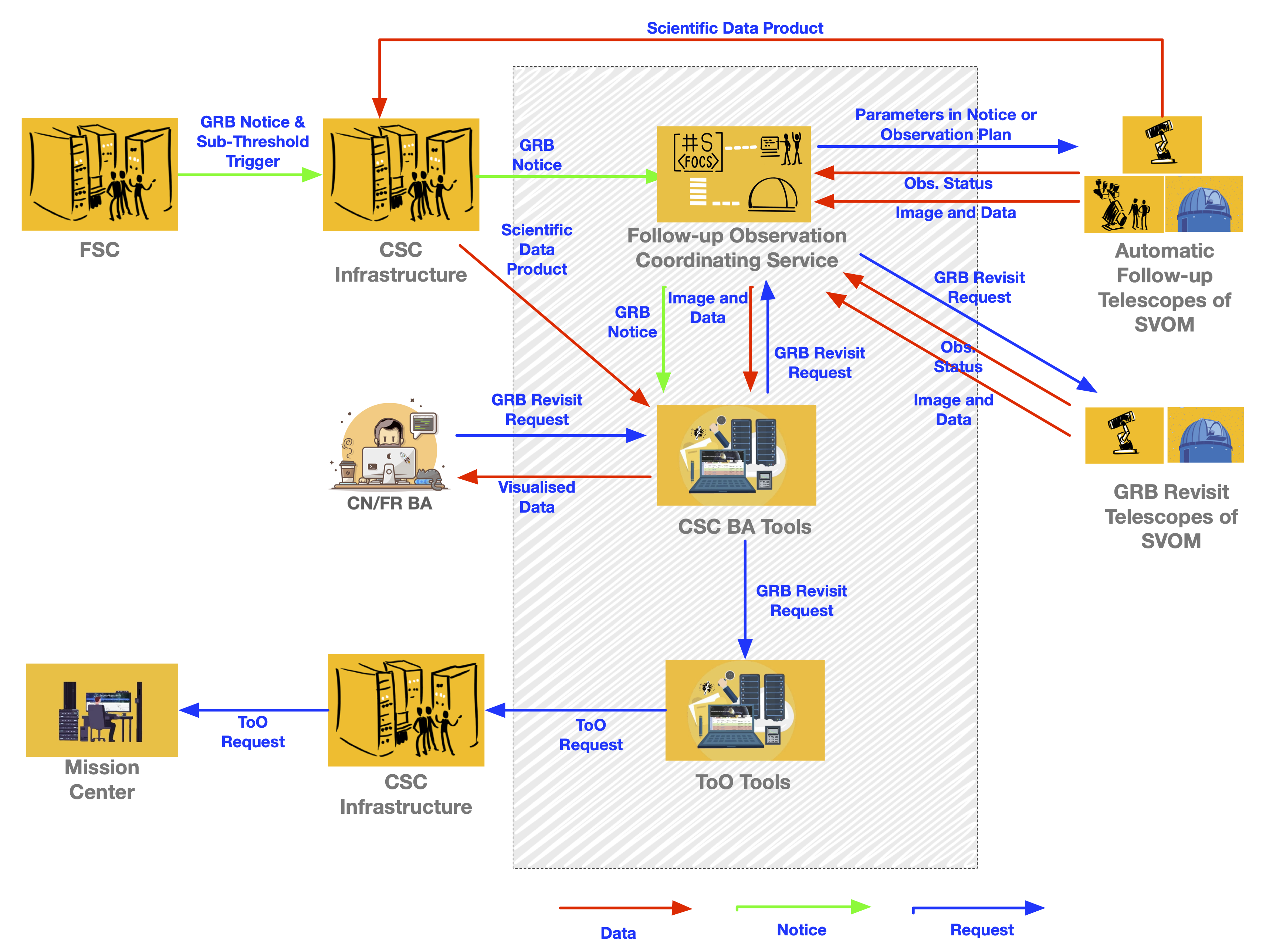}	
	\caption{The operational process for automatic follow-up observations and rapid revisit observations is illustrated in the figure.} 
	\label{fig:followup_observation}
\end{figure*}

The Follow-up Scheduling System implements automatic follow-up observations for GRBs (both space-based and ground-based) and rapid revisit observations for GRB optical counterparts through two technical solutions. When SUSS receives a GRB alert triggered by \textit{SVOM}/Eclairs, it sends the alert information to various clients of the Follow-up Observation Coordinating Service (FOCS, \citep{Han2026}) based on user subscription needs via the FOCS server. Each telescope deploys a user-customized FOCS client program, which generates follow-up observation plans and triggers the telescope to perform automatic follow-up observations. For rapid revisit observations, BAs can generate revisit observation requests through the Follow-up Scheduling webpage in the CSC BA tools and send them to the FOCS server. The FOCS server then forwards them to the FOCS client, which generates observation plans and triggers the telescope to perform rapid revisit observations. Figure \ref{fig:followup_observation} presents a schematic of the operational process for automatic follow-up and rapid revisit observations.

\subsubsection{BA Shift Management System}

The Chinese and French BA teams are responsible for 60\% and 40\% of the duty time, respectively, and manage their schedules independently. From UTC 0h to 13h each day, the CSC BA tools support the Chinese BA team on duty. The BA Shift Management System provides scheduling services for the Chinese BA team; the BA manager can set duty dates, shift handover times, etc., for each BA. The system provides the Chinese duty information to the French iFSC tools (\citep{Louvin2026}) and displays the duty schedules of both the Chinese and French BA teams.

\subsection{ToO Tools}

The \textit{SVOM} mission encompasses three distinct Types of ToO: ToO Nominal (ToO-NOM), ToO Exceptional (ToO-EX), and ToO Multi-Messenger (ToO-MM).

Both ToO-NOM and ToO-EX observations are initiated through user-submitted proposals, which undergo review and approval by a ToO Scientist or the Principal Investigator (PI) before implementation.

In contrast, ToO-MM observations are typically automated: the system generates a list of potential sky regions for \textit{SVOM} to observe based on trigger data from significant transient events, such as gravitational waves or neutrinos. This automatically proposed observing plan is then reviewed by a ToO Scientist or PI before moving into the implementation phase.

The scientific organization of \textit{SVOM}'s ToO observations requires execution by ToO scientific users with the support of ToO Tools. Therefore, the ToO Tools provide the following functionalities:

\subsubsection{Proposal Submission Platform}

\textit{SVOM} users use the Proposal Submission Platform to create and edit ToO-EX and ToO-NOM observation proposals and submit requests through the ToO tools.

The ToO process demands high timeliness, necessitating robust support for both rapid proposal submission and efficient review. The Proposal Submission Platform facilitates this by providing proposal templates, validating target parameters, pre-populating observational configuration combinations, and allowing the import of historical proposals to accelerate preparation.

\subsubsection{Proposal Review Platform}

ToO scientists use the Proposal Review Platform to review ToO-EX and ToO-NOM observation proposals, adjusting and refining observation configuration parameters. The platform streamlines the review process by integrating a target visibility checking feature and implementing mobile-web optimization, enabling ToO scientists to evaluate proposals efficiently from their smartphones.

\subsubsection{ToO Scheduling Dashboard}

ToO Scientists are also responsible for planning ToO observations, determining specific targets, their observational priority, optimal time windows, and exposure durations. They must also promptly adjust existing ToO plans in response to rapid observational demands triggered by significant astronomical events such as GRBs.

The ToO Scheduling Dashboard supports these tasks by enabling scientists to monitor both completed and pending \textit{SVOM} observations. Through visual interactive tools, users can quickly adjust observation plans via simple drag-and-drop operations. The dashboard facilitates the online generation and submission of ToO observation plans (ToO Requests) to the \textit{SVOM} Mission Center (MC, \citep{Liu2026}) and displays the status of ToO execution plan (ToO Workplan) generation.

\begin{figure*}
	\centering 
    \includegraphics[page=1,width=1\linewidth]{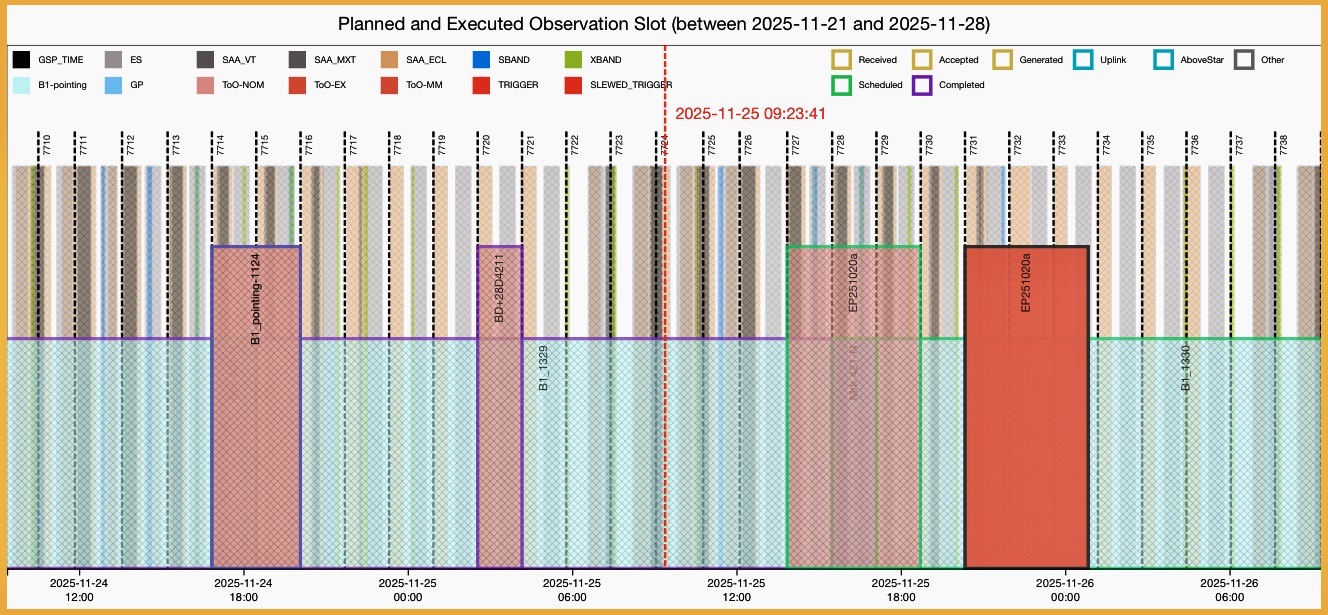}	
	\caption{Schematic of the observation scheduling interface. Distinct rectangles and lines represent different targets, showing their observation status and concurrent satellite states (slewing, eclipse, SAA passage, ground station contact). ToO scientists schedule ToOs by dragging the red boxes to set the time window and inputting the desired duration.} 
	\label{fig:too_scheduler}
\end{figure*}

The diagram (Figure \ref{fig:too_scheduler}) illustrates different observation targets represented by distinct rectangles and lines. For each target, it displays the observation status and satellite operational states during the observation—including satellite slewing maneuvers, entries into eclipse periods, passages through the South Atlantic Anomaly (SAA) region, and contacts with the telemetry download and telecommand upload ground stations\footnote{The communication windows are critical for ToO scientists to predict when observation commands can be uploaded to the satellite and when scientific data will be downlinked, facilitating more effective observation planning.}. ToO scientists can schedule the observation time window for ToOs by dragging the red boxes (targets awaiting observation) left or right, and set the observation duration by entering the required value. 

\subsubsection{ToO Status Display}

The ToO Status Display shows the execution status of ToOs, including request generation, work plans, observation completion status, and data arrival status. ToO scientists can monitor the entire processing and completion status of ToOs.

\subsubsection{ToO Data Product Explorer}

The ToO tools organize scientific data products from space-based instruments and ground telescope detections by observation proposal. ToO scientists and proposers can view the observation data for a specific proposal through the ToO Data Product Explorer. The explorer also provides services to search, view, and retrieve data based on criteria such as target name, coordinates, observation ID, and time range.

\subsubsection{Proposal Alert Module}

The Proposal Alert Module, integrated into the ToO tools, automatically notifies ToO scientists of newly submitted proposals via email and SMS through the Push Notification Service, enabling them to rapidly process ToO requests and reduce observation latency.

\subsection{GP Tools}

The General Program (GP) serves as the primary observational framework for the \textit{SVOM} mission's routine operations. Each GP cycle is typically finalized prior to the start of a calendar year, establishing the specific observation targets and their allocated time slots for the upcoming year.

The GP development process, coordinated through the CSC, involves multiple stages: a "Call for Proposals" (CfP), proposal submission, proposal evaluation, time allocation, and the generation of a GP Catalog. This catalog is subsequently submitted to the \textit{SVOM} MC, where it is incorporated into the comprehensive GP Yearly Planning.

All GP-related activities within the CSC are conducted using the dedicated GP Tools, which provide the following functionalities:

\subsubsection{Proposal Submission Platform}

During the CfP campaign of each GP semester (typically one year), GP proposers use the Proposal Submission Platform to create and submit two types of GP observation proposals: GP-PPT (Pre-Planned Target) for anticipated pre-planned target observations, and GP-ToO for non-planned target observations (i.e., known targets with unpredictable active phases).

The Proposal Submission Platform enables users to complete proposals by providing all required information, including scientific objectives, observation strategy, target lists, and observation configurations. It supports bulk target import and offers auxiliary tools—such as the Target Visibility Calculator, B1 Reference Pointing Calculator, and VT Exposure Tool—to assist proposers in configuring observations accurately.

\subsubsection{Proposal Review and Time Allocation Platform}

All GP proposals (including both GP-PPT and GP-ToO) collected during the CfP campaign undergo review by the GP Manager. The Proposal Review Platform supports this process by providing functionality for evaluating observation proposals and adjusting GP observation configuration parameters.

The Time Allocation Committee (TAC) utilizes the Time Allocation Platform to evaluate observation proposals, determine their acceptance, and allocate observing time and priorities to those approved.

\subsubsection{GP Catalog Generation Platform}

To handle the hundreds or even thousands of GP targets in each observing season, targets from GP proposals are automatically ingested into the system, with their parameters and observational configurations validated automatically. The GP Catalog Generation Platform efficiently manages GP targets, enables GP Managers to select targets in bulk, adjust observational configurations, generate the annual GP target catalog, and supports its submission to the MC to finalize the yearly GP observing plan.

\subsubsection{GP Status Display}

The GP Status Display shows the execution status of GPs, including work plans, observation completion status, and data arrival status. GP managers can monitor the entire processing and completion status of GPs.

\subsubsection{GP Data Product Explorer}

\begin{figure*}
	\centering 
	\includegraphics[page=1,width=1\linewidth]{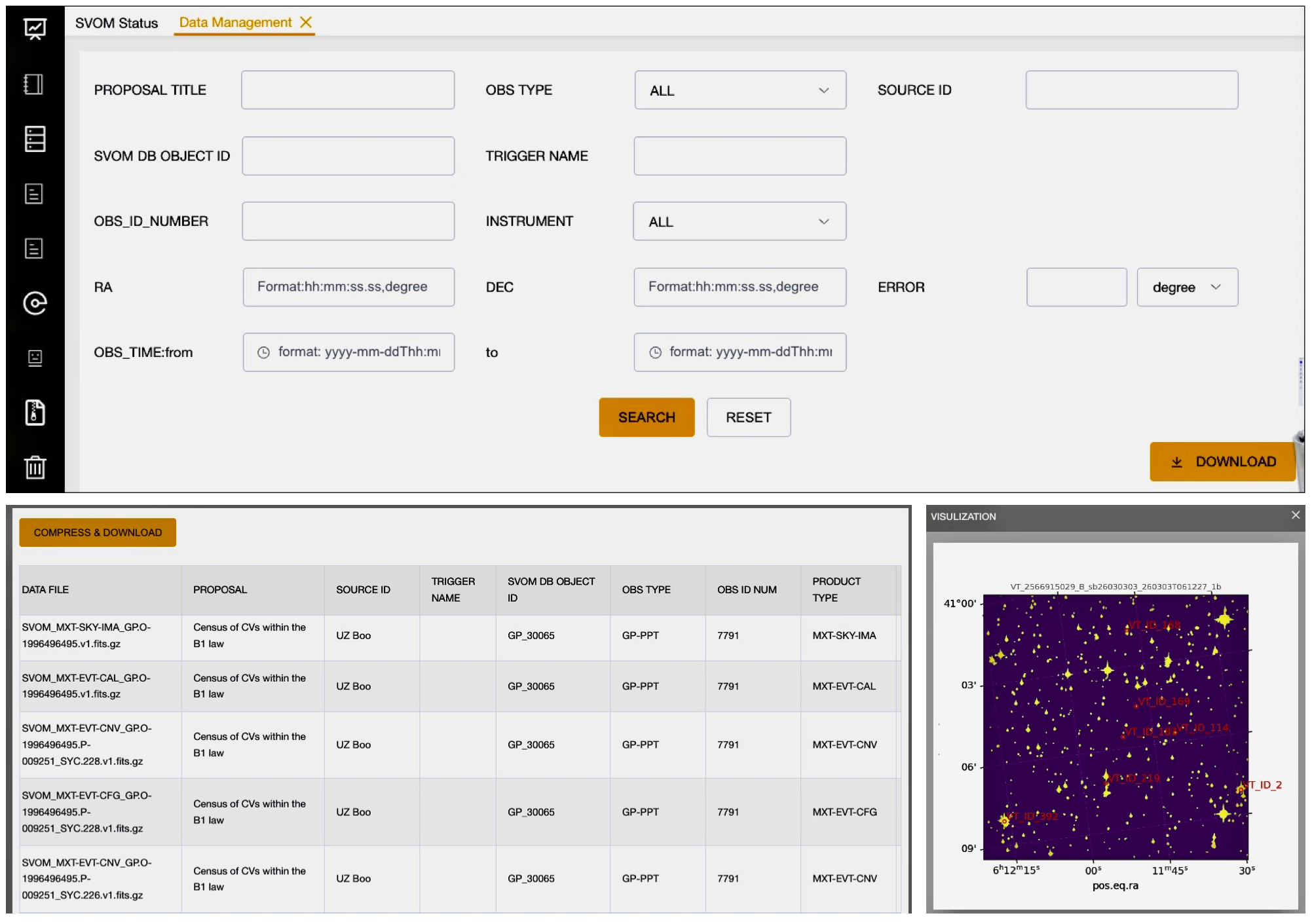}	
	\caption{The GP Data Product Explorer interface. The top panel shows the query interface with search criteria fields (proposal title, target name, coordinates, observation ID, time range). The lower left panel displays the list of retrieved multi-instrument SVOM data products from a sample search. The lower right panel provides an optical image preview of the selected target.} 
	\label{fig:gp_data_product}
\end{figure*}

Although the main instruments used in GP observations are the Microchannel X-ray Telescope (MXT, \citep{Götz2026}) and VT, all four instruments (Eclairs , \citep{Godet2026}, Gamma-Ray Monitor (GRM, \citep{Sun2026}), MXT, VT) are considered to be involved and activated by default. The GP tools are used to organize scientific data products from those instruments by observation proposal. GP managers and proposers can view the observation data for a specific proposal through the GP Data Product Explorer. For example, users can search for all observations of a specific target by entering its name or coordinates, and retrieve the corresponding data products. The explorer also provides services to search, view, and retrieve data based on criteria such as proposal title, target name, coordinates, observation ID, instrument, and time range. Figure \ref{fig:gp_data_product} shows the GP Data Product Explorer interface, including:

\begin{itemize}
\item The data product query interface with search criteria fields

\item A list of retrieved multi-instrument SVOM data products from a sample search

\item An optical image preview panel
\end{itemize}

\subsubsection{Proposal Alert Module}

The Proposal Alert Module automatically notifies GP managers of newly submitted proposals via email through the Push Notification Service.

\section{Arrangement of the Science Operations}
\label{science_operation}

To provide context for \textit{SVOM}'s scientific activities and SUSS service offerings, this chapter presents an overview of their organizational workflow.

\subsection{Science Activities Related with BA}

\subsubsection{BA Shift Scheduling}

The Chinese and French BA teams are managed and scheduled independently. The CSC BA Tools provide management functions for the Chinese BA team. The Chinese BA manager can set the daily duty BA personnel, shift schedules, and handover times. BAs can receive automatically pushed duty information through multiple platforms and view the duty schedules of both teams via the CSC BA tools. Shift handovers can be performed via WeChat groups and Mattermost channels.

\subsubsection{Strategy for Trigger Validation}

The validation of \textit{SVOM} GRB authenticity is completed by the ECLAIRs Trigger Advocate , the GRM instrument team, and the BAs using instrument centers and iFSC tools. BAs can annotate the authenticity of each \textit{SVOM} GRB in the CSC BA Tools.

\subsubsection{GRB Follow-up Organization}

To meet the need for rapid follow-up observations of \textit{SVOM} GRBs, the CSC BA Tools implement rapid follow-up observation request planning and submission in two phases.

First, the automatic follow-up phase generates follow-up observation requests automatically, triggering automatic ToO-NOM-GRB by the \textit{SVOM} satellite (when the automatic slewing threshold is not met) and Chinese ground telescopes (including core telescopes like CGFT and GWAC, and partner telescopes).

Second, the revisit phase involves manually submitted deep photometry or spectroscopy revisit observation requests by BAs. Chinese ground telescopes complete the automatic revisit observation process upon receiving the request. For \textit{SVOM} satellite revisit observations, BAs manually submit ToO observation proposals through the ToO Tools, which handle the creation, planning, and sending of observation requests to the MC.

\subsection{ToO Proposal Handling Procedure}

\begin{figure}
	\centering 
	\includegraphics[page=1,width=1\linewidth]{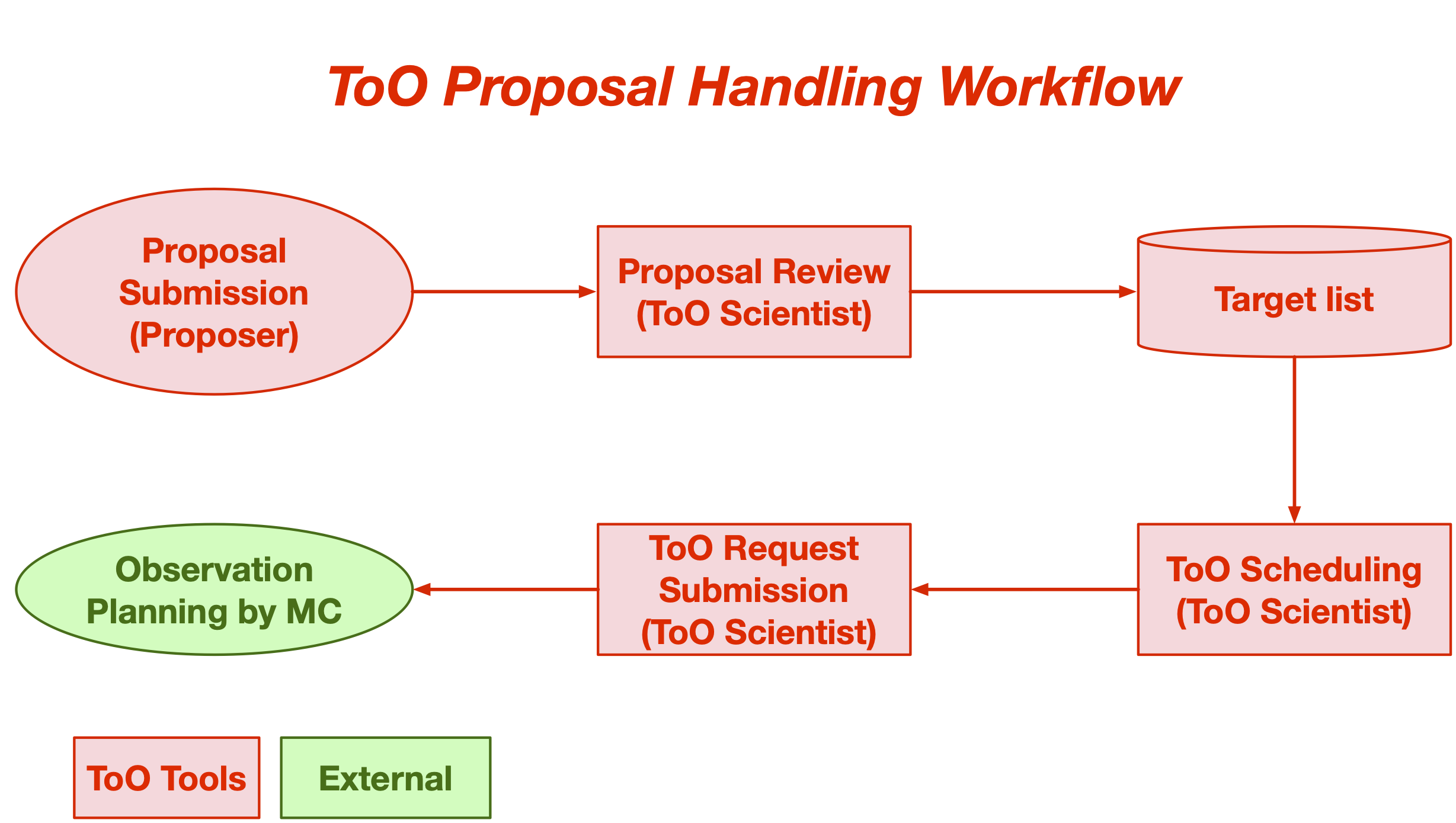}	
	\caption{The figure illustrates the workflow of ToO proposal handling process.} 
	\label{fig:too_workflow}
\end{figure}

ToO offers two observation modes with different priorities: ToO-EX and ToO-NOM. ToO-EX has a higher priority than the GRB observation mode, while ToO-NOM has a priority lower than GRB but higher than the GP observation mode. The handling procedure is the same for both types, as illustrated in Figure \ref{fig:too_workflow}.

Members of the \textit{SVOM} science consortium can submit ToO observation proposals through the ToO Tools. A proposal includes the requested ToO type, scientific objective, parameters of the astronomical target, and observation parameters. Proposers must check target visibility, VT exposure times, and other necessary information before submission.
Each ToO proposal must be reviewed and approved by a ToO scientist before proceeding. ToO scientists evaluate proposals based on scientific importance, target observability, and operational feasibility.

For approved proposals, ToO scientists review the targets and observation parameters, and visually configure the priority, observation duration, and start time for the ToO target, thereby planning the observation tasks and submitting observation requests. When planning, they consider the sequence of different tasks and the observation duration of each task. Typically, the end time of the previous task is selected as the start time for the subsequent task to avoid overlaps. However, for sudden, important tasks, scientists can specify observation requests that override other ongoing ToO tasks.

There are two uplink channels for ToO observation requests at the MC: via the BeiDou link (via the Short Message Service of the Chinese BeiDou Navigation Satellite System) and via the S-band. The typical uplink time delays are different; the BeiDou link has a delay on the order of minutes (usually within 10 minutes), while the S-band has a delay on the order of hours (usually over 8 hours). To execute ToO observation tasks faster, the \textit{SVOM} CSC and MC are working together to optimize the process. By automatically using the GSP sub-satellite point time matching satellite orbital maneuvers as the observation task start time, and automatically temporarily modifying the GSP sub-satellite point time for the next orbital maneuver, the start time of observation tasks can be shortened to within 30 minutes of request submission, significantly enhancing scientific output.

The processing status of each submitted observation proposal can be tracked in the ToO Tools. After submitting an observation request, ToO scientists can view the work plan for the observation task, its execution status, data reception status, and completion status.

\subsection{GP Proposal Handling Procedure}

\begin{figure}
	\centering 
	\includegraphics[page=1,width=1\linewidth]
    {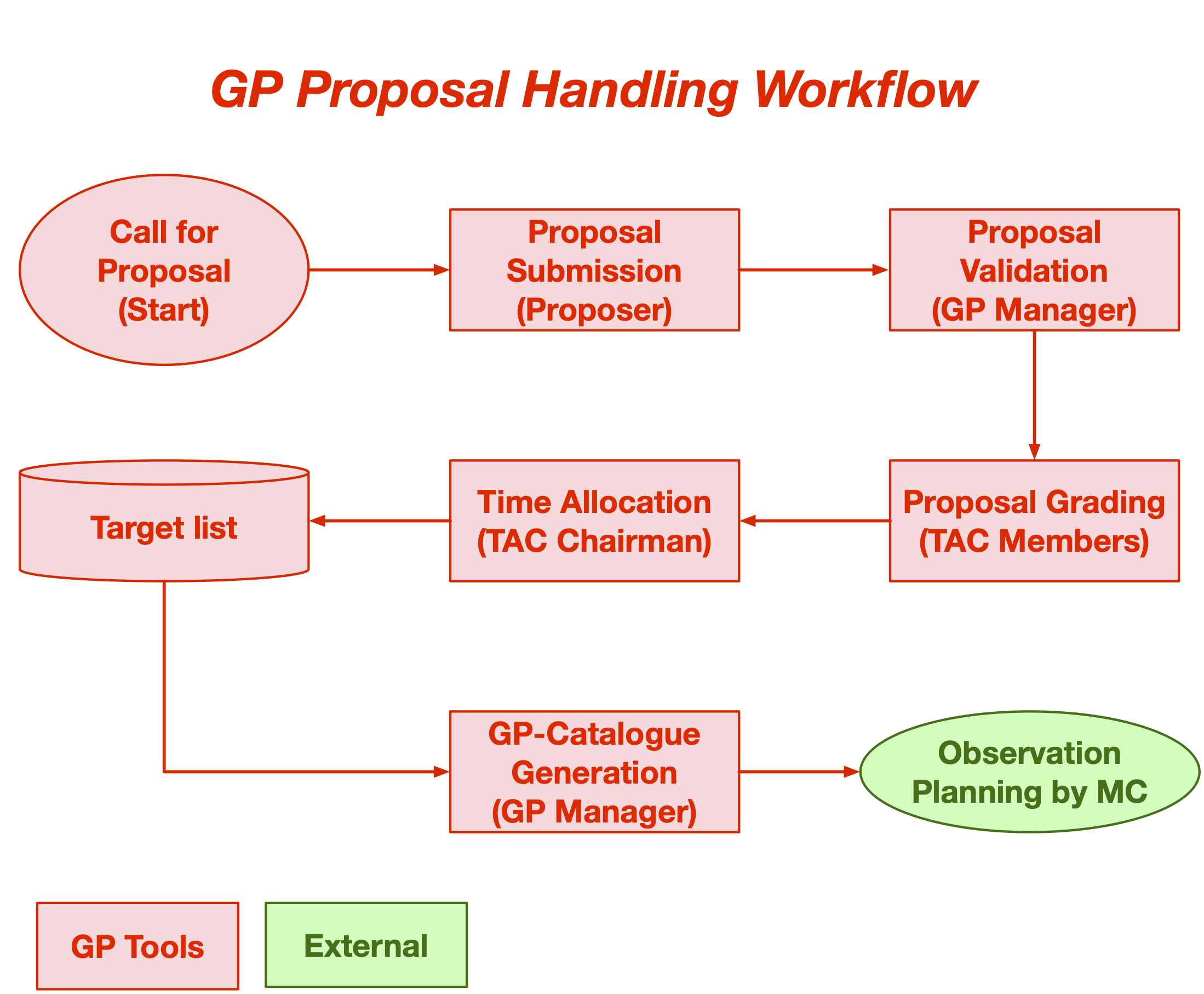}	
	\caption{The figure illustrates the simplified workflow of GP proposal handling process.} 
	\label{fig:gp_workflow}
\end{figure}

\textit{SVOM} solicits GP observation proposals once a year. The GP manager releases the GP CfP announcement via SUSS Portal, initiating the procedure illustrated in Figure \ref{fig:gp_workflow}. The annual proposal call semester typically starts in May, and proposers have 3–4 months to prepare and submit proposals.

The GP proposal call receives two types of proposals: GP-PPT for pre-planned target observations, and GP-ToO for non-planned target observations. GP-PPT proposals are suitable for targets with predictable behavior, such as monitoring known active galactic nuclei (AGNs) or cataclysmic variables. GP-ToO proposals are designed for sources that undergo unpredictable outbursts, such as X-ray binaries or extreme blazer flares, allowing observations to be triggered when the source becomes active.

Both types must be submitted through GP Tools. Proposers must declare the proposal type and its scientific category. The TAC categorizes and reviews proposals based on scientific domain. Currently, GP accepts proposals in areas such as AGNs, TDEs/Clusters/Galaxies, FRBs/Magnetars/AXPs/SGRs, X-ray binaries, ULXs, CVs/Novae, and other galactic and miscellaneous sources.

GP proposals must clearly elaborate on the scientific background, overall significance, relevant references, and prior results. Proposers need to describe the observation plan, state immediate analysis objectives, and elaborate on the proposed targets and specific observation content. For GP-ToO proposals, the observation strategy must clearly provide the trigger criteria and trigger probability.

The GP Manager oversees the validation of each proposal's format and completeness and assigns them to TAC members.

The TAC is responsible for reviewing and evaluating each proposal based on scientific significance, feasibility of the observation strategy, and reasonableness of the observation configuration. The TAC organizes a review meeting each year around October–November to formally approve proposals and allocate the total observation time. It provides recommendations on the observation strategy and configuration for each proposal or target.

The TAC's proposal review, approval, and recommendations are completed through GP Tools. The approved GP observation targets form a master target list.

Based on the TAC's recommendations, the GP manager compiles a one-year GP source catalog (GP-Catalog), including targets and observation parameter configurations. The \textit{SVOM} MC uses this catalog to complete the annual long-term observation planning. The catalog aims to allocate approximately 60\% of the total effective GP observation time to proposals submitted by Chinese Co-Investigators and approximately 40\% to those submitted by French Co-Investigators.

\begin{itemize}
\item Observation time for proposals led by Chinese Co-Investigators is deducted from the Chinese allocation (60\%).

\item Observation time for proposals led by French Co-Investigators is deducted from the French allocation (40\%).
\end{itemize}

During observation execution, GP observation tasks may be interrupted by GRB or ToO tasks. The \textit{SVOM} MC performs re-planning weekly to ensure the total observation time for each GP target. The actual observation time for each target is tracked to strive to maintain the preset 60/40 allocation ratio as much as possible.

The processing status of each submitted observation proposal can be tracked in GP Tools. After submitting the GP source catalog, the GP manager can view the work plans for observation tasks, their execution status, data reception status, and completion status.

GP managers and proposers can query observation tasks and scientific data products using criteria such as proposal name, task type, target name, target ID, coordinates, coordinate range, and observation time range. For instance, a user can search for all observations related to a particular proposal by entering the proposal name, or retrieve data for a specific target by providing its coordinates and a time range. They can also package and download scientific data products.

\begin{figure*}
\centering
\includegraphics[page=1,width=1\linewidth]{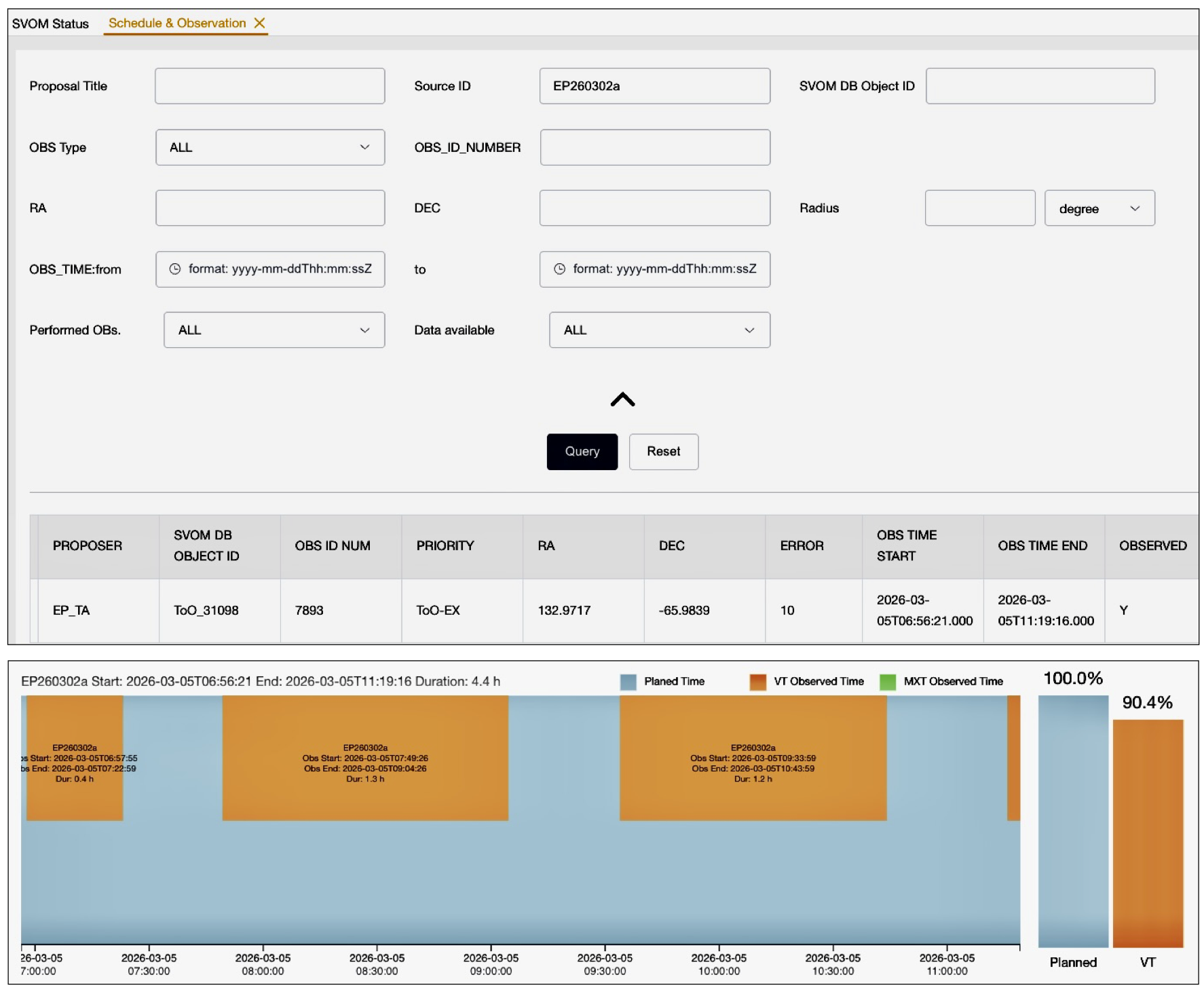}
\caption{The GP observation task and status query interface. The top panel shows the search criteria fields (proposal name, task type, target name, coordinates, time range, etc.). The middle panel lists the retrieved observation tasks with detailed information including proposal name, PI, target, observation type, coordinates, scheduled time, completion status, and data downlink status. The bottom panel provides a visualisation of the observation completion progress.}
\label{fig:gp_query}
\end{figure*}

\section{SUSS Performances in the First Operation Year}
\label{suss_performance}

Since the launch of \textit{SVOM}, the SUSS platform has provided stable service for over 16 months to more than 140 users, including proposers, BAs, GP managers, ToO managers, and TAC members, from more than 30 Chinese and French research institutes and universities, including the NAOC, IHEP, CEA, IAP; GXU, BNU, IRAP etc.\footnote{
National Astronomical Observatories, Chinese Academy of Sciences (NAOC); 
Institute of High Energy Physics, Chinese Academy of Sciences (IHEP); 
Commissariat à l'énergie atomique et aux énergies alternatives (CEA); 
Institut d'Astrophysique de Paris (IAP); 
Guangxi University (GXU); 
Beijing Normal University (BNU); 
Institut de Recherche en Astrophysique et Planétologie (IRAP).} 

The ToO tools have successfully facilitated the processing and observation planning for over 634 ToO proposals (from 2025 January 15 to 2025 November 09). During this period, a series of targeted enhancements were implemented to significantly accelerate the workflow. These included automating GRB follow-up ToO submission, optimizing the proposal handling logic, introducing rapid-processing alerts, deploying a visual ToO scheduling dashboard, and increasing the flexibility of uplink switching conditions for ToO tasks. Collectively, these upgrades have drastically reduced the end-to-end processing time for ToO operations. Prior to these enhancements, ToO-EX proposals typically required up to 12 hours from submission to execution, while ToO-NOM proposals required up to 48 hours. Following the implementation of the upgrades, both ToO-EX and ToO-NOM proposals now consistently achieve processing times of less than 1 hour, markedly improving the mission's responsiveness to transient events.

The GP tools have supported the management of GP targets and the generation of GP catalogs for both the 2024 and 2025 cycles. The GP tools have been instrumental in managing GP targets and generating GP catalogs for both the 2024 and 2025 cycles. To date, the scientific task phase for the first \textit{SVOM} public CfP for the 2026 General Program has been successfully completed. The entire workflow—from the CfP, through their submission, review, and time allocation, to the final generation of the GP catalogs—was seamlessly accomplished with the support of the GP tools. 

Meanwhile, the BA tools have supported the continuous expansion of the ground-based telescope network. So far, over ten telescopes and groups—such as the \textit{SVOM} Chinese Core Telescopes (CGFT, GWAC), the 1.6m Multi-channel Photometric Survey Telescope (Mephisto) of Yunnan University, NAOC's 2.16-meter telescope, the KAIT telescope at Lick Observatory (USA), and Shandong University's 1-meter telescope—have been connected to the FOCS service. Furthermore, the introduction of multi-platform message notifications by the BA tools has significantly improved support for the team's rapid GRB processing workflows.

To continuously meet the evolving needs of scientific users and optimize workflows, SUSS has been updated through three major releases alongside multiple minor iterations.

\section{Conclusion}
\label{conclusion}

The SUSS has been successfully developed and deployed as a cornerstone of the \textit{SVOM} Chinese Ground Segment. It provides a unified, web-based platform integrating the CSC BA Tools, ToO Tools, and GP Tools, effectively centralizing support for the core scientific operations of the \textit{SVOM} mission.

Throughout its first operational year, SUSS has demonstrated robust performance and stability in facilitating the complex workflows associated with Gamma-Ray Burst follow-up (Core Program), ToO observations, and GP management. The system has proven essential for tasks ranging from rapid GRB alert dissemination and BA shift management to streamlined proposal submission, review, scheduling, and data access for ToO and GP programs. Its integrated architecture has enhanced operational efficiency by providing users with a single point of access for information and tools, supported by comprehensive observation monitoring capabilities.

The initial evaluation confirms that SUSS effectively meets its primary objectives and user requirements, serving over 140 users reliably. The insights gained from this initial operation phase, particularly regarding user interaction and system performance under real-world conditions, provide valuable guidance for future development. Ongoing enhancements will focus on further optimizing the user experience, expanding functionality, and improving automation to better serve the evolving needs of the \textit{SVOM} scientific community and maximize the scientific return of the mission.

\begin{acknowledgements}
The Space-based multi-band astronomical Variable Objects Monitor (\textit{SVOM}) is a joint Chinese-French mission led by the Chinese National Space Administration (CNSA), the French Space Agency (CNES), and the Chinese Academy of Sciences (CAS). We gratefully acknowledge the unwavering support of NSSC, IAMCAS, XIOPM, NAOC, IHEP, CNES, CEA, and CNRS.
The authors are also thankful for support from the National Key R\&D Program of China (grant No.2024YFA1611700, 2024YFA1611702, 2023YFA1608304). This work is also supported by the Strategic Priority Research Program of the Chinese Academy of Sciences (grant No. XDB0550101, XDB0550401). 
We thank the staffs of the \textit{SVOM} Mission Center for important assistants during development and deployment of the SUSS system.
\end{acknowledgements}

\bibliographystyle{raa}
\bibliography{ms2026-0015}

@ARTICLE{Liu2026,
   author = {{Liu}, Y. and {Bai}, M. and {Wei}, M. and et al.},
    title = "{SVOM Mission Center Operations and Coordination}",
  journal = {Research in Astronomy and Astrophysics},
     year = 2026,
    volume = {to appear},
    pages = {--},
}

@ARTICLE{Huang2026,
   author = {{Huang}, M. and {Zheng}, S. and et al.},
    title = "{SVOM Chinese Ground Segment Infrastructure and Data Processing}",
  journal = {Research in Astronomy and Astrophysics},
     year = 2026,
    volume = {to appear},
    pages = {--},
}

@ARTICLE{Cordier2026,
   author = {{Cordier}, B. and {Wei}, J. and {Zhang}, S. and et al.},
    title = "{The SVOM Mission: Scientific Objectives and Operations}",
  journal = {Research in Astronomy and Astrophysics},
     year = 2026,
    volume = {to appear},
    pages = {--},
}

@ARTICLE{Gehrels04,
   author = {{Gehrels}, N. and {Chincarini}, G. and {Giommi}, P. and et al.},
    title = "{The Swift Gamma-Ray Burst Mission}",
  journal = {The Astrophysical Journal},
     year = 2004,
    volume = 611,
    pages = {1005-1020},
      doi = {10.1086/422091},
}

@ARTICLE{Atwood09,
   author = {{Atwood}, W.~B. and {Abdo}, A. and {Ackermann}, M. and et al.},
    title = "{The Large Area Telescope on the Fermi Gamma-ray Space Telescope Mission}",
  journal = {The Astrophysical Journal},
     year = 2009,
    volume = 697,
    pages = {1071-1102},
      doi = {10.1088/0004-637X/697/2/1071},
}

@ARTICLE{Yuan22,
   author = {{Yuan}, W. and {Zhang}, C. and {Chen}, Y. and et al.},
    title = "{The Einstein Probe Mission}",
  journal = {Handbook of X-ray and Gamma-ray Astrophysics},
     year = 2022,
    pages = {86},
      doi = {10.1007/978-981-16-4544-0_151-1},
}

@ARTICLE{Han2021,
   author = {{Han}, X. and {Xiao}, Y. and {Zhang}, P. and {Turpin}, D. and {Xin}, L. and {Wu}, C. and et al.},
    title = "{The Automatic Observation Management System of the GWAC Network. I. System Architecture and Workflow,}",
  journal = {Publications of the Astronomical Society of the Pacific,},
     year = 2021,
    volume = 133,
    issue = 1024,
    pages = {17},
      doi = {10.1088/1674-4527/21/9/215},
}

@ARTICLE{Xin2026,
   author = {{Xin}, L. and {Huang}, L. and {Cai}, H. and et al.},
    title = "{GWAC Operations and Data Processing for SVOM}",
  journal = {Research in Astronomy and Astrophysics},
     year = 2026,
    volume = {to appear},
    pages = {--},
}

@ARTICLE{Qiu2026,
   author = {{Qiu}, Y. and {Wang}, J. and {Ho}, L. and et al.},
    title = "{The SVOM Visible Telescope: Instrument Design and Performance}",
  journal = {Research in Astronomy and Astrophysics},
     year = 2026,
    volume = {to appear},
    pages = {--},
}

@ARTICLE{Wu2026,
   author = {{Wu}, C. and {Kang}, Z. and {Lu}, X.~M. and et al.},
    title = "{Chinese Ground Follow-up Telescope System for SVOM}",
  journal = {Research in Astronomy and Astrophysics},
     year = 2026,
    volume = {to appear},
    pages = {--},
}

@ARTICLE{Han2026,
   author = {{Han}, X. and {Zhang}, P. and {Xiao}, Y. and {Zhang}, R. and {Wu}, C. and {Xin}, L. and et al.},
    title = "{SVOM Follow-up Observation Coordinating Service}",
  journal = {arXiv:2511.06647},
     year = 2026,
    volume = {to appear},
    pages = {--},
}

@ARTICLE{Louvin2026,
   author = {{Louvin}, D. and {Corre}, D. and {Formica}, A. and et al.},
    title = "{French Burst Advocate Support Tools for SVOM}",
  journal = {Research in Astronomy and Astrophysics},
     year = 2026,
    volume = {to appear},
    pages = {--},
}

@ARTICLE{Götz2026,
   author = {{Götz}, D. and {Mercier}, K. and {Schanne}, S. and et al.},
    title = "{The SVOM Microchannel X-ray Telescope: Instrument Design and Performance}",
  journal = {Research in Astronomy and Astrophysics},
     year = 2026,
    volume = {to appear},
    pages = {--},
}

@ARTICLE{Godet2026,
   author = {{Godet}, O. and {Cordier}, B. and {Atteia}, J.~L. and et al.},
    title = "{The SVOM ECLAIRs Telescope: Gamma-ray Burst Detection and Localization}",
  journal = {Research in Astronomy and Astrophysics},
     year = 2026,
    volume = {to appear},
    pages = {--},
}

@ARTICLE{Sun2026,
   author = {{Sun}, X. and {Li}, T. and {Zhang}, S. and et al.},
    title = "{The SVOM Gamma-Ray Monitor: Instrument Design and Performance}",
  journal = {Research in Astronomy and Astrophysics},
     year = 2026,
    volume = {to appear},
    pages = {--},
}

\label{lastpage}

\end{document}